\documentclass[preprint2]{aastex}
\usepackage[]{natbib}
\begin{document}
\title{Evidence that an Intrinsic Redshift Component that is a Harmonic of z = 0.062 May be Present in Every Quasar Redshift}
\author{M.B. Bell\altaffilmark{1}}
\altaffiltext{1}{Herzberg Institute of Astrophysics,
National Research Council of Canada, 100 Sussex Drive, Ottawa,
ON, Canada K1A 0R6}

\begin{abstract}

After estimating and removing all ejection-related Doppler components from the redshifts of the QSOs near NGC 1068, the remaining redshift is assumed to be intrinsic. This well-studied case is the first example in which it has finally been possible to separate the intrinsic redshift component from the cosmological and other Doppler components. It is shown that this leads to intrinsic redshift components that occur at exact multiples of z = 0.062 and are defined by the relation z$_{\rm i}$ = 0.062[10$N - M]$ where $N$ and $M$ can have only certain discrete values.
It is also shown that the intrinsic redshifts given by this relation
agree closely with the redshifts found to be preferred in analyses of quasar emission-line redshifts. It is now immediately apparent that the peaks in the distribution of quasar emission line redshifts also occur at pure harmonics of z = 0.062. Although these peaks have previously been fitted by a log(1+z) relation, the above explanation now seems more appropriate. These results indicate that the $\Delta$z = 0.062 redshift interval, long known to be present in QSO emission-line redshifts below z = 0.6, is also present in an intrinsic component of QSO redshifts up to z $\sim$ 2. This appears to imply that the $\Delta$z = 0.062 redshift interval represents a fundamental redshift spacing that arises from the very nature of the intrinsic redshifts.

\end{abstract}
\keywords{galaxies: individual (NGC 1068) -- galaxies: Seyfert --quasars:general}

\section{Introduction}

It has been shown \citep{bel02a,bel02b} from the redshifts and positions of 12 of the compact, high-redshift objects near NGC 1068 that they appear to have been ejected outwards along the rotation axis of NGC 1068 in four similarly structured triplets that differ only in their size, rotation angle and rotation axis tip angle. In the previous studies \citep{bel02a,bel02b} no attempt was made to remove the redshift components due to the triplet ejection velocities. Assuming here that the model is correct, it is shown that it is possible to determine the rotation axis tip angle for each triplet at the time of its ejection. All ejection-related Doppler components can then be removed allowing the residual intrinsic redshift components to be obtained. This is the first instance where an attempt has been made to derive accurate values for the intrinsic components. Previously, they have always been contaminated with cosmological and other Doppler-related components. It is shown further that the intrinsic redshift values derived for the sources near NGC 1068 can be used to define a general relation, and that this relation, in turn, predicts the values found to be preferred in emission line redshift distributions covering redshifts from z = 0.06 to z = 2.

\begin{figure}
\hspace{-1.0cm}
\vspace{-2.0cm}
\epsscale{1.0}
\plotone{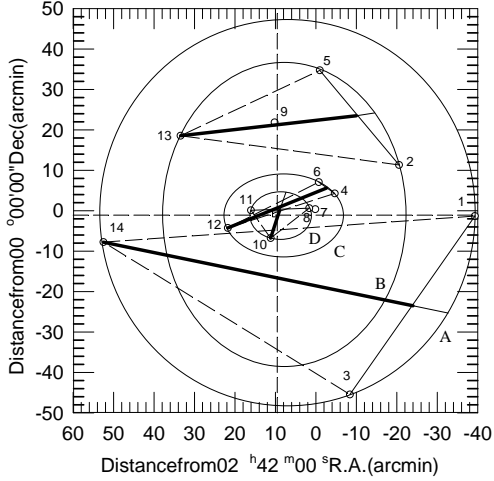}
\caption{\scriptsize{Source triplets near NGC 1068 are defined by the four ellipses labelled A - D. Objects designated as pairs are joined by a solid line. Associated singles are connected by dashed lines. The location of NGC 1068 is shown by the open square. The rotation axis of the central torus is indicated by the vertical dashed line through the galaxy. Single and pair midpoints are joined by heavy solid lines. \label{fig1}}}
\end{figure}

\begin{figure}
\hspace{-1.0cm}
\vspace{-2.0cm}
\epsscale{1.0}
\plotone{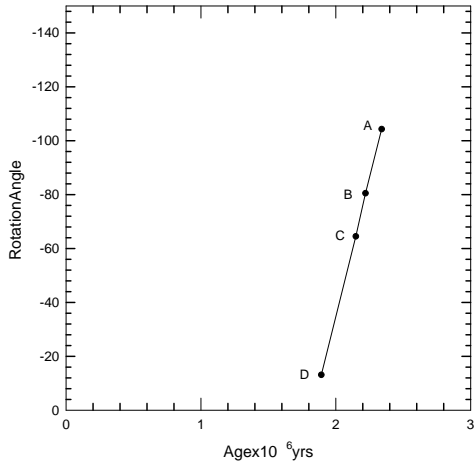}\caption{\scriptsize{Age versus rotation angle. \label{fig2}}}
\end{figure}

\section{Calculating Intrinsic Redshift Components}

The compact objects near NGC 1068 \citep{bur99} are plotted in Fig. 1, where the heavy solid lines connect the single and paired source in each triplet, and the rotation axis of the central torus is indicated by the vertical dashed line. The four triplets are defined by the four ellipses labeled A to D.
In the ejection model presented earlier \citep{bel02a,bel02b}, each triplet is ejected outward along the rotation axis of the central torus. The axis is assumed to have an associated tip angle $\gamma$ defined as its angle of tip toward the observer at the top. The triplet center of mass (CofM) moves outward along this axis and its speed and direction must be known if the Doppler components resulting from this motion are to be calculated and removed. Because the tip angle was unknown, no attempt was made to remove this ejection component in the previous studies. In addition to this motion, the triplets further divide into a single and a pair that separate with equal velocities in opposing directions, each perpendicular to the direction of triplet ejection. The position angle defined by the line joining the single and pair-midpoint (the heavy lines in Fig. 1) gives the rotation angle ($\theta$) of the relevant triplet (where $\theta$ is measured counter-clockwise from North). It was shown previously \citep{bel02a} how $\theta$ can be calculated if the rotation axis tip angle is known.

a) Estimating the tip angle $\gamma$.

Previously the tip angle for all four triplets was assumed to be $\gamma = 18\arcdeg$. However, there appears to be strong evidence to indicate that it may be significantly larger than this for all triplets. Since the orbital plane is perpendicular to the rotation axis, if the triplet CofM is ejected along the rotation axis, then it is likely that the triplet plane will remain perpendicular to the direction of ejection. The plane defined by the pair and singlet in each triplet thus defines the orientation of the rotation axis (the tip angle) at the time the triplet was ejected. 

In this scenario, if $\gamma$ = 0 for a given triplet, its three sources would all lie along an E-W (horizontal) line in Fig. 1. In triplets A and B, where $\theta$ is close to -90$\arcdeg$, any increase in $\gamma$ above 0 introduces a N-S separation on the sky in the relevant paired sources. For triplets A and B the tip angles are given by the angle $\beta$, defined as the angle between the horizontal and the line joining the two sources making up the pair. In triplet D, where $\theta$ is close to 0$\arcdeg$, an increase in $\gamma$ would introduce a N-S displacement in the relevant pair-single separation. Thus, for triplet D the tip angle is given approximately by the angle between the horizontal and the line joining the single and pair midpoint. In the case of triplet C, whose rotation angle lies closer to -45$\arcdeg$, the above approach is less accurate.

The $\gamma$ values obtained using the above technique for triplets A, B, C and D are listed in col 7 of Table 1. These values are all significantly larger than the value $\gamma = 18\arcdeg$ assumed previously, and in closer agreement with the tip angle near $60\arcdeg$ currently measured for the optical axis of NGC 1068 \citep{sch00,boc00}. The tip angle for triplet A can be either $55\arcdeg$ or $125\arcdeg$. The correct value appears to be $125\arcdeg$. The ejection model requires this when the relative orientations of the red- and blue-shifted components making up the various pairs are taken into account. This conclusion also means that triplet A will have been ejected in the same jet direction as the other three triplets and not in the counter-jet direction as previously assumed. The CofM of triplet A will then move toward us instead of away, and this must be taken into account when the Doppler components are removed.

b) Calculating Doppler Components.

It is assumed here that the tip angles calculated above, which have large uncertainties ($\sim 5\arcdeg-10\arcdeg$), can at least act as a rough indication of the true tip angle. It is further assumed that other physical relations, which must be satisfied by the ejection model, can be used to define the tip angle more accurately. For example, if the model is valid, the event ages derived should vary smoothly and, if the mass is constant, linearly with the rotation angle. 

In order to examine how changes in the tip angle affect the age of each triplet, a spreadsheet was prepared that calculated new ages and intrinsic redshifts with the insertion of each new value of $\gamma$. Since the redshifts of sources 10 and 12 have not been measured it was also necessary to examine a range of redshifts for these sources. This too was accommodated in the spreadsheet calculations. Initial values were obtained by assuming that the pair-single ejection velocities of triplets C and D were similar to those measured for triplets A and B. Column 9 of Table 1 list the redshifts derived for sources 10 and 12. Table 2 lists various stages in the calculation of triplet ages and Table 3 summarizes the stages involved in calculating intrinsic redshifts. 

The $\gamma$-values listed in column 2 of Table 2 and the redshifts listed for the C and D triplets in column 5 were arrived at using the values in col 7 of Table 1 as a guide and requiring the $\theta$ vs Age relation to be linear.

c) Rotation period

The resulting $\theta$ vs Age relation is shown in Fig. 2. The slope of this line gives a rotation period for the nucleus of NGC 1068 of $1.8\times10^{6}$ yr, which is a factor of 3.3 shorter than that found by \citet{all01} for the central torus. This value is assumed to be more accurate than that given earlier \citep{bel02a} since the earlier age calculations did not include accurate corrections for all ejection velocities. Since the age is calculated from the present position of the objects and their ejection velocity perpendicular to the l-o-s, this would have affected the previous age calculations.

\begin{deluxetable}{ccccccccc}
\tabletypesize{\scriptsize}
\tablecaption{Triplet Sources with relevant
distances, estimated tip angle and redshifts. \label{tbl-1}}
\tablewidth{0pt}
\tablehead{
\colhead{Triplet(sources)} & \colhead{Pair} &  \colhead{Pair Mean Dist. \tablenotemark{a}} &  \colhead{Singl.}  &  \colhead{Singl. Dist.\tablenotemark{a}} & \colhead{Tripl. Dist.\tablenotemark{a}}  & \colhead{$\gamma(\arcdeg$)} & \colhead{z$_{\rm mean}$} & \colhead{z$_{\rm singl}$}
}
\startdata
A(1-3,14) & 1-3  & 48.85   &   14  & 43.9 & 46.4 & 55(125) & 0.4935 & 0.655 \\
B(2-5,13) & 2-5  & 35.05   &   13  & 30.4 & 32.7 &  49   & 0.761 & 0.684 \\
C(4-6,12) & 4-6  & 14.4    &   12  & 12.1 & 13.3 & $>36$ & 1.1005 & 1.041\tablenotemark{b} \\
D(8-11,10) & 8-11 & 7.30   &  10   &  6.0  & 6.65 &  73  &  1.2015 &  1.068\tablenotemark{b} \\
\enddata 
\tablenotetext{a}{distances represent angular distance on the sky from NGC 1068
in arcmin.}
\tablenotetext{b}{fitted using age vs $\theta$ relation}
\end{deluxetable}

\begin{deluxetable}{cccccccccc}
\tabletypesize{\scriptsize}
\tablecaption{Tip Angle($\gamma$), Rotation Angle ($\theta$), Ejection Velocity, and Age of Triplets. \label{tbl-2}}
\tablewidth{0pt}
\tablehead{
\colhead{Trip.} & \colhead{($\gamma$)} & \colhead{$\theta$ (deg)\tablenotemark{a}} & \colhead{(90-$\theta$)} & \colhead{z$_{\rm p-s\|}$\tablenotemark{b}} & \colhead{z$_{\bot}$\tablenotemark{c}}  & \colhead{v$_{\bot}(\times10^{4}$)\tablenotemark{d}} & \colhead{z$_{\rm eject}$\tablenotemark{e}} & \colhead{s-p Displ.\tablenotemark{f}} & \colhead{Age ($\times10^{6}$)\tablenotemark{g}}
}
\startdata
A & 124.8 & -104.2 & 14.3 & $\mp0.0513$ & -0.355 & -8.85 & -0.366 & 43.5 & 2.34 \\
B & 46.3 & -80.4 & 9.5 & $\pm0.0223$ & 0.191 & 5.20 & .194 & 24.3 & 2.22  \\
C & 73.1 & -64.4  & 25.5 & $\pm0.0144$ & 0.103 & 2.94 & 0.114 & 13.3 & 2.15 \\
D & 80.3 & -13.0 & 76.9 & $\pm0.0313$ & 0.0428 & 1.26 & 0.190 & 5.0 & 1.89 \\

\enddata
\tablenotetext{a}{calculated as described by \citet{bel02a}.}
\tablenotetext{b}{pair-single l-o-s ejection component (from col 8 and 9 of Table 1.)}
\tablenotetext{c}{z$_{\bot}$   = z$_{\rm p-s\|}$/[tan($90-\theta$)x(cos$\gamma$)]}
\tablenotetext{d}{Uses z$_{\bot}$ and eqn 9.11 of \citet{bur67} to obtain velocity in km s$^{-1}$.}
\tablenotetext{e}{z$_{\rm eject}$ = z$_{\rm p-s\|}$/[sin($90-\theta$)x(cos$\gamma$)]}
\tablenotetext{f}{mean angular separation from NGC 1068 of pair and single on the sky.}
\tablenotetext{g}{Age(yr) = 0.476 x (col 9/col 7); assumes 1$\arcmin$ = 5 kpc at distance of NGC 1068.}
\end{deluxetable}

\begin{deluxetable}{cccccccc}
\tabletypesize{\scriptsize}
\tablecaption{Calculation of CofM ejection velocity and Intrinsic Redshifts. \label{tbl-3}}
\tablewidth{0pt}
\tablehead{
\colhead{Triplet} & \colhead{vert.displ\tablenotemark{a}} & 
\colhead{v$_{\rm trip\bot}(\times10^{4})$\tablenotemark{b}} &  \colhead{z$_{\rm trip\bot}$}   & \colhead{z$_{\rm trip\|}$\tablenotemark{c}}  &  
\colhead{z$_{\rm mean}$} & \colhead{z$_{\rm T}$\tablenotemark{d}}   & 
\colhead{z$_{\rm i}$\tablenotemark{e}}
}
\startdata
A  & 15.2  & 3.092 & 0.1090 & -0.1568  &  0.4935 &  0.5745  & 0.867 \\
B  & 20   &  4.285 & 0.1547  & -0.1619 & 0.761   &  0.7225  & 1.055   \\
C  & 2   &  4.436 & .0149  &  -0.0490 & 1.1005   &  1.0707  & 1.1775   \\
D  & 1  & 0.251  & 0.0084  & -0.0491  & 1.2015  & 1.1348  & 1.2451  \\
\enddata
\tablenotetext{a}{vertical displacement on the sky of triplet CofM}
\tablenotetext{b}{Triplet CofM velocity perpendicular to the l-o-s in km s$^{-1}$ = (0.476$\times$ col 2)/age}
\tablenotetext{c}{(z$_{\rm trip\bot})\times$(tan$\gamma$) = l-o-s Doppler component of CofM.}
\tablenotetext{d}{(1 + z$_{\rm T}$) = (1 + z$_{\rm mean}$)/(1 + z$_{\|}$). Z$_{\rm T}$ is total CofM redshift, (i.e. intrinsic plus ejection velocity.)}
\tablenotetext{e}{ assumes (1+z$_{\rm i}$) = (1+z$_{\rm T}$)/(1 + z$_{\rm trip\|}$)}
\end{deluxetable}

\section{Intrinsic Redshift Relation}

\begin{figure}
\hspace{-1.0cm}
\vspace{-1.0cm}
\epsscale{1.0}
\plotone{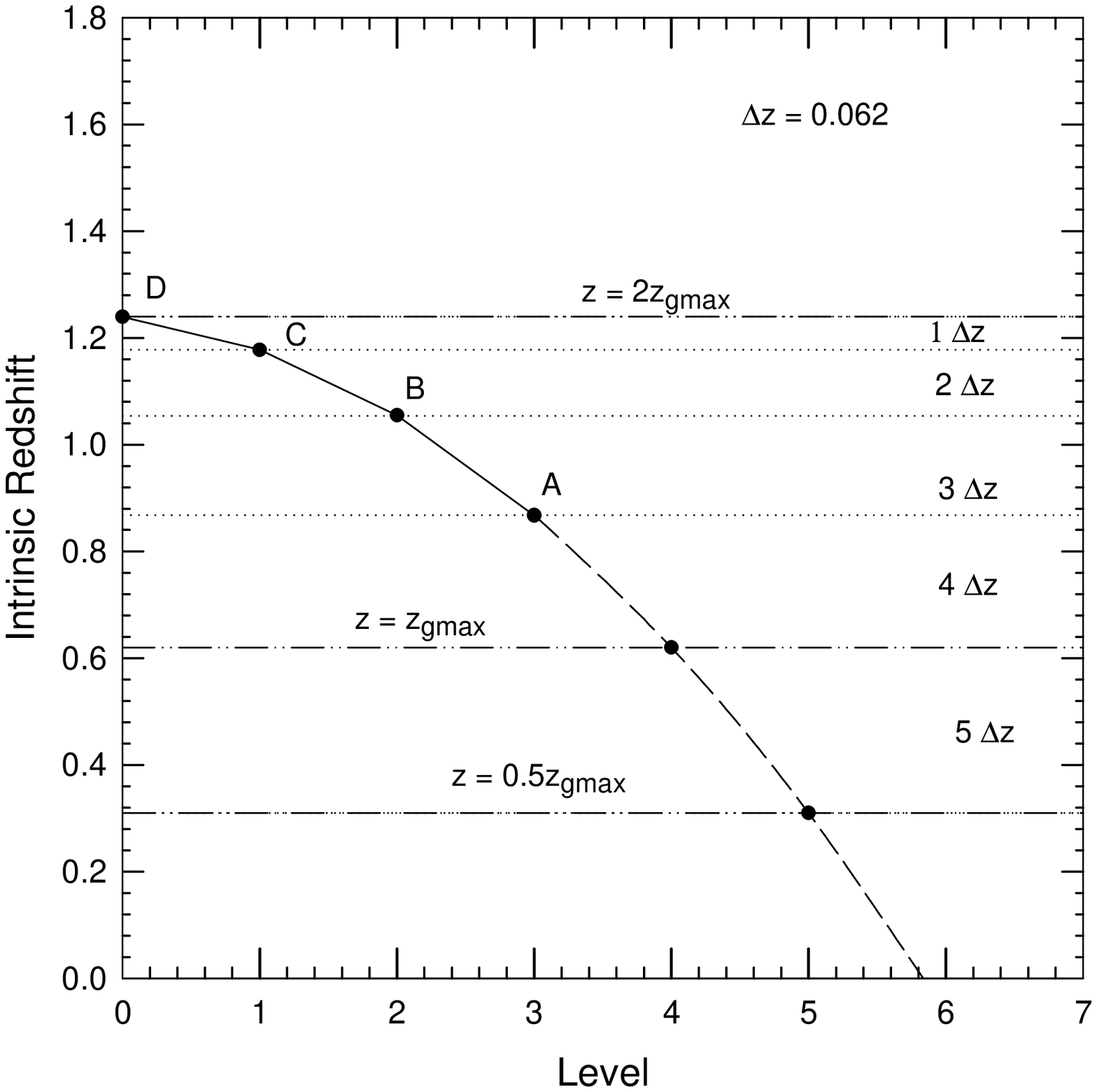}\caption{\scriptsize{Intrinsic redshift vs level for the N = 2 state. See text for an explanation of z$_{\rm gmax}$. \label{fig3}}}
\end{figure}

The intrinsic redshifts calculated in Table 3 are plotted versus level (triplet) in Fig. 3 where \em the redshift spacing between levels can be seen to increase in multiples of the redshift increment $\Delta z$ = 0.062. \em Before the Doppler components due to triplet ejection were removed here, the redshifts (z$_{\rm mean}$) were found to exhibit similar characteristics \citep{bel02a}, except in that case the redshift increments between levels were found to be multiples of $\Delta z$ = 0.05. As long as the ejection velocities are small, and the rotation axis is tipped toward the observer, the slightly larger increment of $\Delta z$ = 0.062 found here after removal of the Doppler components would be anticipated. It is therefore not surprising that a similar relation was visible in the raw data. This also reinforces the likelihood that the results obtained here have not been introduced by the above calculations. The most interesting new result is that when all Doppler components are accounted for, the intrinsic redshifts are all multiples of the redshift increment $\Delta$ z = 0.062 first found by Burbidge over thirty years ago \citep{bur68,bur90}.

The intrinsic redshift values found for the four triplets near NGC 1068 are labeled A, B, C, and D in Fig. 3. 
Two additional intrinsic redshift values (0.62 and 0.31) can be predicted from Fig. 3 corresponding to levels 4 and 5. These appear to be the only other acceptable values since higher ones would be blueshifted. 

These intrinsic redshift components can be defined by the relation:

z$_{\rm i} = 0.062[10N - M]$ ----------------- (1)

where $N$ = 2, $M_{2} = [n(n+1)/2]$, and $n$ = 0, 1, 2, 3, 4, and 5.

For N = 2, this relation predicts intrinsic redshift values of 0.31, 0.62, 0.868, 1.054, 1.178, and 1.24, all separated by multiples of the redshift increment $\Delta z$ = 0.062, and all occurring at harmonics of z = 0.062. Thus this increment is present in the intrinsic component for redshifts well above the z = 0.06 to 0.6 redshift range considered by \citet{bur90}. 

\section{Preferred Emission-Line Redshift Values found in Previous QSO Redshift Analyses}

Several investigators have carried out searches for periodicities in quasar emission-line redshift distributions. It is important to note that all of these were carried out using observed redshifts that can include large Doppler ejection and cosmological components in addition to an intrinsic component. More than thirty years ago \citet{bur68} pointed out that there were strong peaks near z = 0.061 and 1.95. \citet{bur90} showed that the z = 0.06 periodicity held up when a larger sample (z $\leq$ 0.6) was used. More extensive analyses of the raw redshift distributions have found other preferred redshift values. \citet{kar71,kar77,bur78,bur01} reported that the redshift distributions showed peaks near z = 0.061, 0.30, 0.60, 0.96, 1.41, 1.96, 2.36, 3.45 and 4.47. \citet{kar77} pointed out that these redshift peaks can be reasonably well represented by the relation $\Delta$log(1+z) = 0.089.

The distribution found for QSO emission line redshifts below z = 1.24 \citep{kar77} is reproduced in Fig. 4. This plot has been chosen because it represents one of the first investigations. Although several follow-up investigations have been reported \citep{arp90,bur01}, the peaks below z = 2.0 remain relatively unchanged. These later investigations also represent cases where data have been selected after a relation has been previously found and this can sometimes lead to data choices that reinforce the original relation. No such bias can be present in the initial investigation. 

In Fig. 4 the heavy vertical bars near the bottom of the figure indicate the locations of the peaks claimed to be periodic in log(1+z). Also included in the figure are vertical dotted lines indicating the locations of the intrinsic redshifts defined by eqn 1 ($N$ = 2) for the sources near NGC 1068 (hereafter referred to as z$_{N = 2}$ values). They range from the maximum value z = 1.24, down to z = 0.31 as $n$ varies from 0 to 5. All occur at pure harmonics of z = 0.062. It is clear from this figure that, even though the emission-line redshifts do not represent a purely intrinsic redshift distribution, there is excellent agreement between the peaks in the distribution and the intrinsic redshifts found from the QSOs near NGC 1068. In fact, for z $\leq 1.24$, the fit to the z$_{N=2}$ redshifts is significantly better than the fit to the log(1+z) periodicity. This implies that the Doppler components present in the redshifts of the sources near each peak must be small. This, in turn, implies that they must be associated with reasonably nearby objects (z$_{c} < 0.02$), and have been expelled approximately perpendicular to the l-o-s. Those not expelled perpendicular to the l-o-s are assumed to be represented by those sources located below the curved baseline in Fig. 5.

In previous analyses \citep{kar77,arp90,bur01} it seems not to have been realized that the emission-line redshift peaks are located at harmonics of z = 0.062. Presumably this can be explained by the facts that the peaks are not periodic in z and that the redshifts used are observed values and not purely intrinsic ones, which may affect the peak positions slightly. However, it is worth noting also that the log(1+z) periodicity has nothing to support it, whereas the z$_{N=2}$ redshifts have been predicted here using an analysis that is completely independent from the redshift distribution analysis. Furthermore, the log(1+z) relation has been known for 25 years and has not led to any other related discoveries. This and the fact that it is based on a distribution that can contain large Doppler components are arguments to question its appropriateness to correctly represent the intrinsic redshift relation.

\section{Other Values of $N$}

The objects clustered near z  = 0.061 appear to be of a different class (many are Seyferts, N-gals, etc.) and are not included in the distribution in Fig. 4. On the other hand, the objects examined by \citet{bur90}, which had redshifts up to z = 0.6, are distributed at single intervals of $\Delta$z = 0.062. One might then speculate that these objects (with a maximum redshift of z = 0.62) represent the $N$ = 1 state of eqn 1. If so, for z$_{N = 1}$, from the results of \citet{bur90}, it can be concluded that $M_{1} = n$ where $n$ can have the ten values 0,1,2....9. Eqn 1 can then be expressed in the form z$_{\rm i} = [0.62 - 0.062n$].

In Fig. 5 all 574 objects examined by \cite{kar77} have been plotted.
The objects with measured redshifts above z = 1.24 may simply be z$_{N = 2}$ sources with large Doppler components that have yet to be removed, or, they may represent the $N$ = 3 state (see discussion below on gravitational redshifts for further comments on redshifts above z = 2).

We have thus seen from the results of \citet{bur90}, that for the $N=1$ state, $M_{1}$ = n. From the NGC 1068 results for the $N=2$ state, $M_{2}$ = $n(n+1)/2$. Thus as $N$ changes from $N$ = 1 to $N$ = 2, $M$ goes from $M_{1} = n$ to $M_{2} = n(n+1)/2$.
If we assume that the $N$ = 3 state behavior can be derived from the $N$ = 2 state in the same manner that the $N$ = 2 state came from the $N$ = 1 state, then for $N$ = 3, M$_{3} = [p(p+1)/2]$, where p = $[n(n+1)/2]$. It can easily be shown that acceptable values of $M$ are then $M_{3}$ = 0,1,6, and 21 where $n$ = 0,1,2, and 3.  This gives intrinsic redshift values for the $N$ = 3 state of 1.86, 1.798, 1.488, and 0.558. As can be seen in Fig. 5 \em these values show excellent agreement with all peaks below z = 2 that were not previously fitted by the z$_{N=2}$ lines. \em This can be seen clearly in Fig. 6 where the curved baseline in Fig. 5 has been removed. The curved baseline, as postulated above, is assumed to represent those sources in which l-o-s Doppler components are sufficiently large to smear out the periodicity in the intrinsic components.

Eqn 1 can thus define all three $N$-states provided the appropriate values for $M$ are used in each case. For $N$ = 1, $M_{1}$ = 0,1,2,3,...9; For $N$ = 2, $M_{2}$ = 0,1,3,6,10,15; for $N$ = 3, $M_{3}$ = 0,1,6, and 21. These values are obtained directly from the relations $M_{1}$ = n; $M_{2} = [n(n+1)/2]$; $M_{3} = [p(p+1)/2]$ (where p = $n(n+1)/2)$. In all cases $n$ = 0, 1,2,3...,n. The redshifts predicted by these values of $N$ and $M$ all occur at harmonics of z = 0.062 and they show the same distribution in redshift as do the emission-line peaks.

\section{Intrinsic Redshift vs $M$}

In Fig. 7 the intrinsic redshift vs $M$ relation is compared for all three N-states. As can be seen, and can be predicted from eqn. 1, the intrinsic redshifts fall off vs $M$ with the slope -0.062 for all three states. This slope is thus given by the periodicity first found by \citet{bur68}.

\begin{figure}
\hspace{-1.0cm}
\vspace{-2.0cm}
\epsscale{1.1}
\plotone{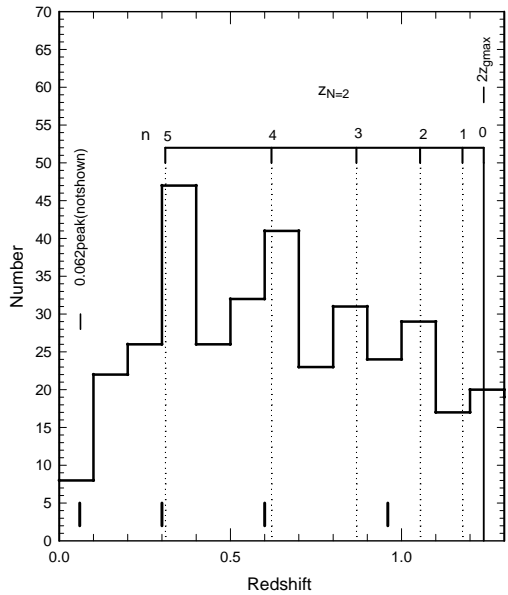}\caption{\scriptsize{Distribution of QSO redshifts from \citet{kar77} for z $< 1.24$. The vertical dotted lines indicate the location of the intrinsic redshifts found here (eqn 1). The short dashes near the bottom indicate the positions of the redshifts defined by the relation reported by \citet{kar77}. See text for the definition of z$_{\rm gmax}$.  \label{fig4}}}
\end{figure}

\begin{figure}
\hspace{-1.0cm}
\vspace{-2.5cm}
\epsscale{1.1}
\plotone{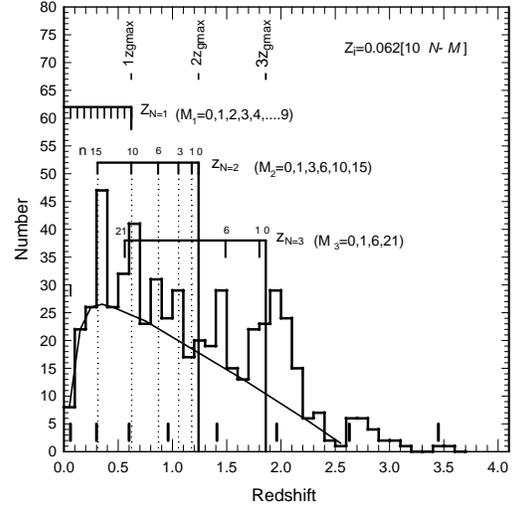}\caption{\scriptsize{Distribution of 574 QSO redshifts from \citet{kar77}. The vertical dotted lines indicate the location of the intrinsic redshifts found here. The short dashes near the bottom indicate the positions of the redshifts defined by the relation reported by \citet{kar77}. See text for a description of the redshifts above 1.86 and the definition of z$_{\rm gmax}$.  \label{fig5}}}
\end{figure}

\begin{figure}
\hspace{-1.0cm}
\vspace{-2.5cm}
\epsscale{1.1}
\plotone{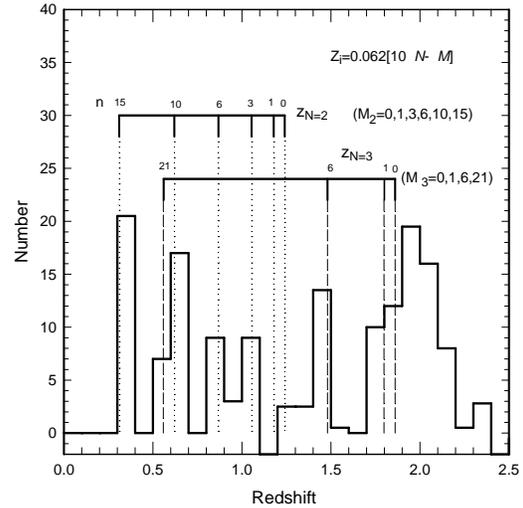}\caption{\scriptsize{Distribution of QSOs with redshifts z $\leq 2.5$ from \citet{kar77} after subtraction of the baseline in Fig. 5. The vertical dotted lines indicate the location of the intrinsic redshifts found here. The vertical dashed lines indicate the location of the intrinsic redshifts predicted for the $N = 3$ state. \label{fig6}}}
\end{figure}

\begin{figure}
\hspace{-1.0cm}
\vspace{-1.0cm}
\epsscale{1.0}
\plotone{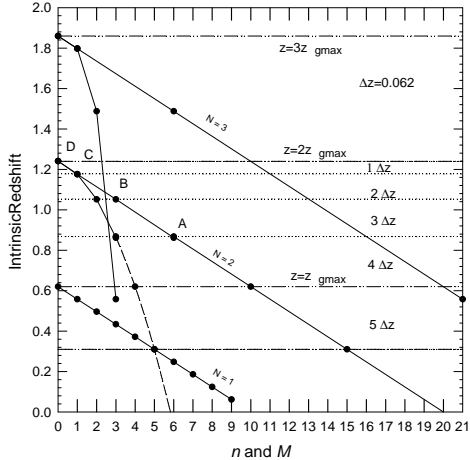}\caption{\scriptsize{Intrinsic redshift vs $n$ (curves) and $M$ (straight lines) for the N = 1, 2, and 3  states. Note that for the $N$ = 1 state the curve and straight line are superimposed. \label{fig7}}}
\end{figure}

\begin{figure}
\hspace{-1.0cm}
\vspace{-2.0cm}
\epsscale{1.0}
\plotone{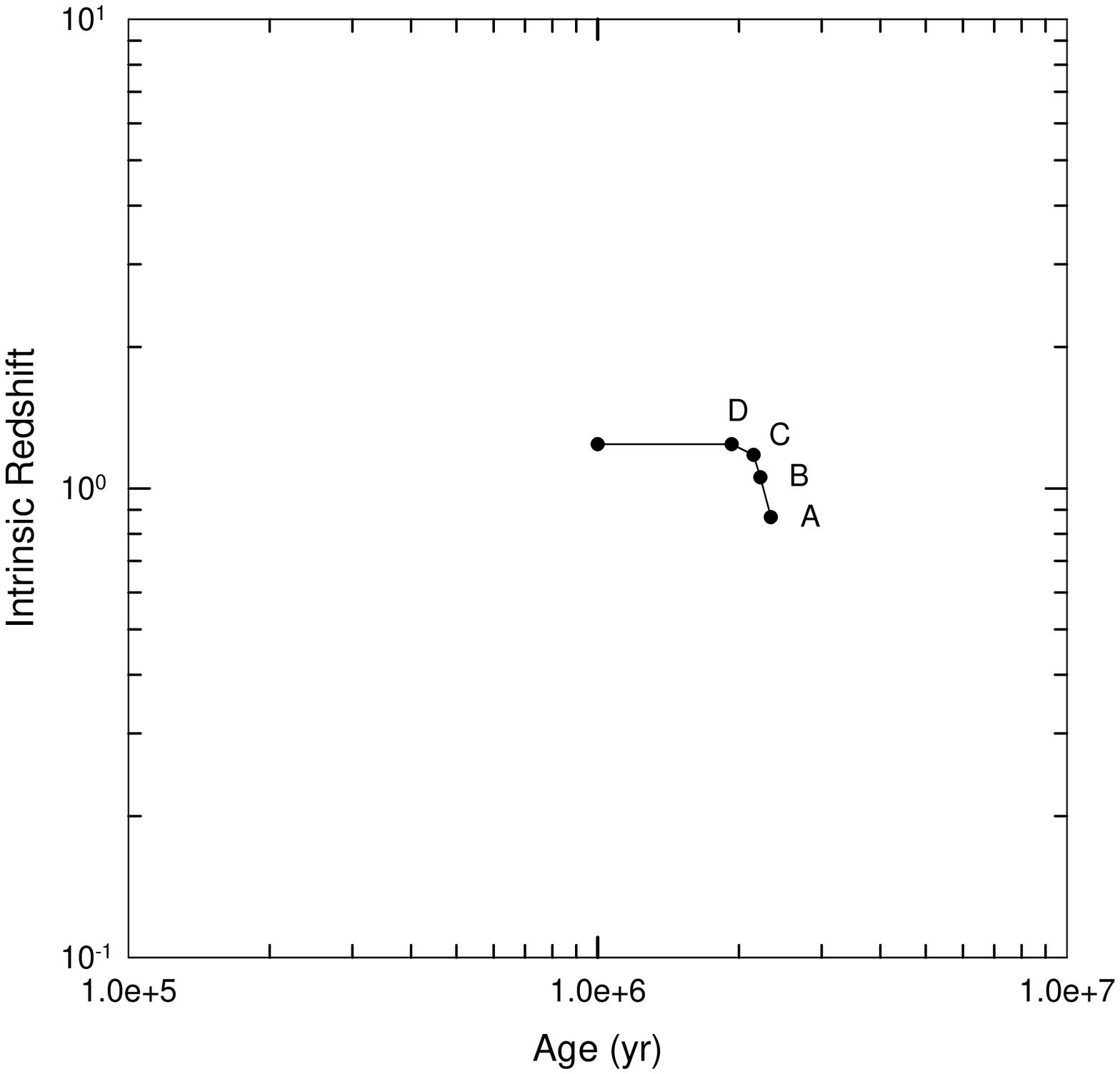}\caption{\scriptsize{Intrinsic redshift vs Age plotted on a logarithmic scale. In the oldest triplets (A and B) the intrinsic redshift components fall off with a slope of -4 and have disappeared by the age of $3\times10^{6}$ yr. \label{fig8}}}
\end{figure}

\section{An Alternate Way of Expressing the Intrinsic Redshift Relation}


Although eqn 1 may represent the simplest form of the intrinsic redshift relation, it is worth pointing out that it can also be expressed in the form z$_{\rm i}$ = [0.62$N$ + 0.062$M$]. Here the constant $10\times0.062$ has been expressed as 0.62. This number is identical to the theoretical maximum gravitational redshift (z$_{\rm gmax}$) obtained from the Schwartzschild interior solution for a perfect fluid sphere \citep{bur67,bon64,buc59,buc66}, and may suggest that the intrinsic redshift component is somehow produced gravitationally. This suggestion is also reinforced by the fact that there appears to be no correlation between the peaks reported above z = 2 and intrinsic redshifts that can be predicted for $N$-values above 3. If the intrinsic redshifts are of gravitational origin this would be consistent with the well-known z = 2 upper limit for gravitational redshifts \citep{bur67}. It is therefore tentatively assumed here that observed redshifts above z = 2 contain large Doppler components but their intrinsic components remain below z$_{\rm i}$ = 1.86.

\section{The Significance of the parameters $N$ and $M$}

At the moment one can only speculate on the significance of the parameters $N$ and $M$. Since $N$ is constant for a given source, its value may be related to some parameter (e.g. mass) of the parent object or its nucleus. If so, this would seem to imply that the nuclear mass is quantized with at least three acceptable levels. Since the ejected objects all have similar masses initially and appear to expel mass (or energy) with time from the central QSO into a surrounding host, the discrete nature of the parameter $M$ may then suggest that this mass is ejected in discrete "packets". 

As far as the formation of galaxies is concerned, these results do not change significantly the picture discussed earlier \citep{bel02a,bel02b}. They still suggest that galaxies are created continuously throughout the life of the Universe. However, if the intrinsic component is gravitationally produced, it would seem less likely that the model of \citet{nar80} plays any role. In Fig. 8 the z$_{N = 2}$ intrinsic redshifts are plotted vs Age on a logarithmic scale. It is clear that these intrinsic components play a very short-lived part in the life of a galaxy, lasting only for the first 3 million years.

\section{Conclusions}

By removing all ejection-related Doppler components from the redshifts of the compact QSOs near NGC 1068 it has been shown that the remaining intrinsic components can be defined by the relation z$_{\rm i}$ = $0.062[10N - M]$ where $N$ = 2, $M = [n(n+1)/2]$ and $n$ = 0,1,2,3,4,5. This is the first instance in which it has been possible to separate out the intrinsic component.
It is shown that this same relation will fit the results obtained by \citet{bur90} for redshifts between z = 0.06 and z = 0.6, if $N$ = 1 and $M = n$, where $n$ = 0,1,2,... 9. The $M$-values for the $N$ = 3 state follow directly from these two states and it is shown that together these three states predict intrinsic redshifts that agree closely with those values found to be preferred in emission-line redshift analyses from z = 0.06 to z = 2. These results indicate that the $\Delta$z = 0.062 redshift interval may thus represent a fundamental redshift spacing that arises from the very nature of the intrinsic redshifts.

Because this relation includes a constant (z$_{\rm gmax}$ = 0.62) which can be related to gravitational redshifts, the intrinsic components may have a gravitational origin. If so, this increment may be evidence for some form of gravitational quantization.

It is concluded here that the log(1+z) relation, previously fitted to the peaks in the emission-line redshift distribution, may not be the appropriate one to describe the intrinsic redshifts.

By fine-tuning the tip angles it has also been possible to predict accurate redshift values for sources 10 and 12. 
It is important to note that values of the Doppler components calculated here could change if the redshifts of sources 10 and 12 can be measured and turn out to be significantly different from those estimated above. Small differences might still be accommodated by adjusting the relevant tip angle, provided the age versus rotation angle relation in Fig. 2 is not destroyed. However, the excellent agreement between the intrinsic redshifts given by eqn 1, and the redshift peaks found in quasar emission-line analyses, strongly suggests that the above relation is a significant one.

\end{document}